\documentclass[12pt,english]{article}
\pdfoutput=1
\usepackage{lmodern}
\DeclareMathAlphabet{\mathbbold}{U}{bbold}{m}{n} 

\usepackage[T1]{fontenc}
\usepackage[latin9]{inputenc}
\usepackage{geometry}
\usepackage{color}
\usepackage{cite}
\usepackage{babel}
\usepackage{amsmath}
\usepackage{amssymb}
\usepackage{cancel}
\usepackage{graphicx}
\usepackage{esint}
\usepackage[unicode=true,pdfusetitle,
 bookmarks=true,bookmarksnumbered=false,bookmarksopen=false,
 breaklinks=false,pdfborder={0 0 1},backref=false,colorlinks=true]
 {hyperref}
\hypersetup{
 citecolor=blue,linkcolor=black,urlcolor=blue}

\makeatletter

 \@ifundefined{textcolor}{}
 {%
   \definecolor{BLACK}{gray}{0}
   \definecolor{WHITE}{gray}{1}
   \definecolor{RED}{rgb}{1,0,0}
   \definecolor{GREEN}{rgb}{0,1,0}
   \definecolor{BLUE}{rgb}{0,0,1}
   \definecolor{CYAN}{cmyk}{1,0,0,0}
   \definecolor{MAGENTA}{cmyk}{0,1,0,0}
   \definecolor{YELLOW}{cmyk}{0,0,1,0}
 }

\usepackage{babel}

\makeatletter
\def\simgt{\mathrel{\lower2.5pt\vbox{\lineskip=0pt\baselineskip=0pt
           \hbox{$>$}\hbox{$\sim$}}}}
\def\simlt{\mathrel{\lower2.5pt\vbox{\lineskip=0pt\baselineskip=0pt
           \hbox{$<$}\hbox{$\sim$}}}}
\makeatother

\newcommand{\be}{\begin{equation}}
\newcommand{\bea}{\begin{eqnarray}}
\newcommand{\ee}{\end{equation}}
\newcommand{\eea}{\end{eqnarray}}
\newcommand\beq{\begin{equation}}
\newcommand\eeq{\end{equation}}

\newcommand{\Eq}[1]{Eq.~(\ref{#1})}
\newcommand{\Eqs}[2]{Eqs.~(\ref{#1}) and (\ref{#2})}

\newcommand{\Fig}[1]{Fig.~\ref{#1}}
\newcommand{\Ref}[1]{Ref.~\cite{#1}}

\newcommand{\ket}[1]{|#1\rangle}
\newcommand{\bra}[1]{\langle #1|}

\begin{document}
\interfootnotelinepenalty=10000
\baselineskip=18pt
\hfill CALT-TH-2015-048
\hfill

\vspace{1.5cm}
\thispagestyle{empty}
\begin{center}
{\LARGE\bf
Wormhole and Entanglement (Non-)Detection in the ER=EPR Correspondence
}

%}\\
\bigskip\vspace{1.5cm}{
{\large
Ning Bao,$^{1,2}$ Jason Pollack,$^{2}$ and Grant N. Remmen$^2$}
} \\[7mm]
{\it
$^1$Institute for Quantum Information and Matter\\ and \\
$^2$Walter Burke Institute for Theoretical Physics,\\[-1mm] California Institute of Technology, Pasadena, CA 91125, USA}

\let\thefootnote\relax\footnote{e-mail: \url{ningbao@theory.caltech.edu},
\url{jpollack@theory.caltech.edu},
\url{gremmen@theory.caltech.edu}} \\
 \end{center}
\bigskip
\centerline{\large\bf Abstract}

\begin{quote}\small
The recently proposed ER=EPR correspondence postulates the existence of wormholes (Einstein-Rosen bridges) between entangled states (such as EPR pairs). Entanglement is famously known to be unobservable in quantum mechanics, in that there exists no observable (or, equivalently, projector) that can accurately pick out whether a generic state is entangled. Many features of the geometry of spacetime, however, are observables, so one might worry that the presence or absence of a wormhole could identify an entangled state in ER=EPR, violating quantum mechanics, specifically, the property of state-independence of observables. In this note, we establish that this cannot occur: there is no measurement in general relativity that unambiguously detects the presence of a generic wormhole geometry. This statement is the ER=EPR dual of the undetectability of entanglement.
\end{quote}

\setcounter{footnote}{0}

\newpage
\tableofcontents

\section{Introduction}

Black holes are the paradigmatic example of a system where both field-theoretic and gravitational considerations are important. Black hole thermodynamics and the area theorem \cite{BHLaws,Bekenstein} already provided a relationship between entanglement and geometry, while the classic black hole information paradox \cite{HawkingInfo} and its potential resolution via complementarity \cite{Complementarity,'tHooft:1984re,'tHooft:1990fr} pointed at the subtlety of the needed quantum mechanical description. In the last few years, the firewall paradox \cite{AMPS} has heightened the tension between these two descriptions, prompting a number of proposals to modify the standard picture to a greater or lesser extent.

One set of proposals \cite{Papadodimas:2012aq,Papadodimas:2013wnh,Papadodimas:2013jku,Papadodimas:2015xma,Papadodimas:2015jra} modifies quantum mechanics to allow for state-dependence of the black hole horizon, so that an infalling observer does not encounter a firewall even though the state of the black hole horizon can be written as a superposition of basis states that each have high energy excitations \cite{Bousso:2013wia}. In order to avoid this problem, the presence or absence of a black hole firewall must become a nonlinear observable, contrary to standard quantum mechanics. In a recent paper, Marolf and Polchinski \cite{MP} have pointed out that this nonlinearity cannot be ``hidden''; if it is strong enough to remove a firewall from a generic state, it must lead to violations of the Born Rule visible from outside the horizon.

In this paper, we consider a different idea inspired by the firewall paradox, the ER=EPR correspondence \cite{ER=EPR}, which asserts the existence of an exact duality between Einstein-Podolsky-Rosen (EPR) pairs, i.e., entangled qubits, and Einstein-Rosen (ER) bridges \cite{ER,Kruskal,Fuller}, i.e., nontraversable wormholes. This duality is supposed to be contained within quantum gravity, which is in itself meant to be a \textit{bona fide} quantum mechanical theory in the standard sense. The ER=EPR proposal is radical, but it is not obviously excluded by either theory or observation, and indeed has passed a number of nontrivial checks \cite{JensenKarch,Sonner,Baez:2014bka,Gharibyan:2013aha,Mandal:2014wfa}; if true, it has the potential to relate previously unconnected statements about entanglement and general relativity in a manner reminiscent of the AdS/CFT correspondence \cite{AdSCFT,Witten,MAGOO}. In a previous paper \cite{BPR}, we pointed out that in ER=EPR the no-cloning theorem is dual to the general relativistic no-go theorem for topology change \cite{GerochThesis,Tipler}; violation on either side of the duality, given an ER bridge (two-sided black hole), would lead to causality violation and wormhole traversability.

In light of the result of \Ref{MP}, one might be worried that ER=EPR is in danger. It is well-known that entanglement is not an observable, in the sense we will make precise below; we cannot look at two spins and determine whether they are in an arbitrary, unspecified entangled state with one another. Yet ER=EPR implies that the two spins are connected by a wormhole, so that the geometry of spacetime differs according to whether or not they are entangled. If this difference in geometry could be observed, entanglement would become a (necessarily nonlinear) observable as well and the laws of quantum mechanics would be violated, contrary to the assumption that the latter are obeyed by quantum gravity.

In this paper, we show that ER=EPR does not have this issue. Unlike the modifications to quantum mechanics considered by \Ref{MP}, wormhole geometry \textit{can} be hidden. In particular, we show that in general relativity no measurement can detect whether the interior of a black hole has a wormhole geometry. More precisely, observers can check for the presence or absence of specific ER bridge configurations, but there is no projection operator (i.e., observable) onto the entire family of wormhole geometries, just as (and, in ER=EPR, for the same reason that) there is no projection operator onto the family of entangled states. 

The remainder of this paper is organized as follows. We first review the basic quantum mechanical statement that entanglement is not an observable. Next we introduce the maximally extended AdS-Schwarzschild geometry in general relativity and, using AdS/CFT, on the CFT side. As a warmup, we first show that no single observer can detect the presence of a wormhole geometry. We then turn to more complicated multiple-observer setups and show, as desired, that they are unable to detect the presence of nontrivial topology in complete generality.

\section{Entanglement is Not an Observable}

The proof that one cannot project onto a basis of entangled states \cite{nielsen2010quantum} proceeds as follows. Assume the existence of a complete basis set of entangled states $\ket{\psi_{E_i}}$, distinct from the basis set of all states. A projection onto this basis could be written in the form
\beq
\hat{P}_E=\sum_i\ket{\psi_{E_i}}\bra{\psi_{E_i}}.
\eeq
Note, however, that the set of all entangled states has support over the entire Hilbert space, as the entangled states can be written as linear sums of unentangled states:
\beq
\ket{\psi_{E_i}}=\sum_{j\in B_i} \ket{\psi_j}
\eeq
for some set $B_i$.
Therefore, the projector onto the set of all entangled states does not project out any states in the Hilbert space. Said another way, the set of all entangled states is not a set that is closed under superposition, thus preventing a projection thereupon. Since no projector exits, entanglement is therefore not an observable.

\section{Setup}

\begin{figure}[t]
  \begin{center}
    \includegraphics[width=0.3\textwidth]{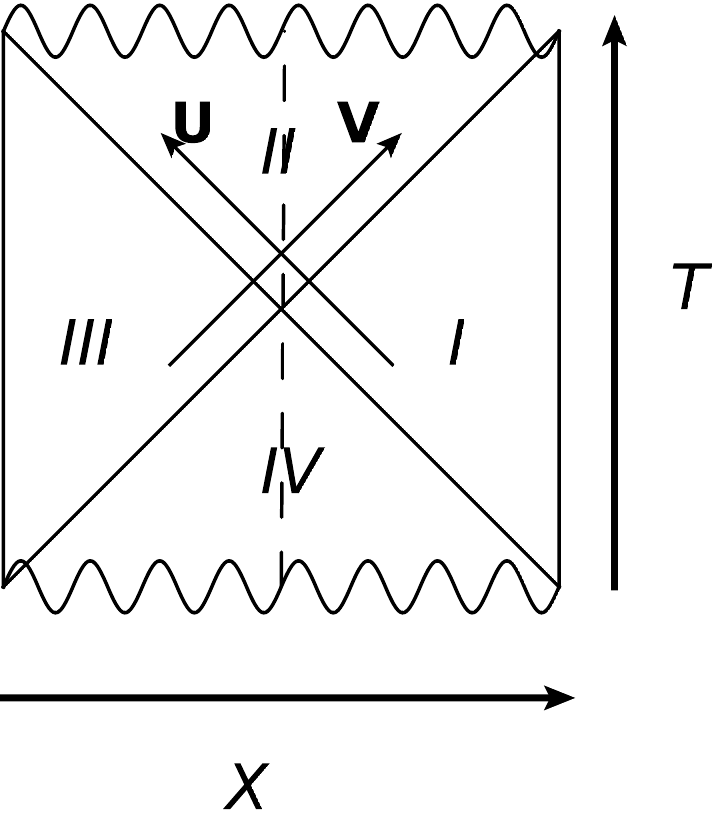}
  \end{center}
  \caption{The maximally extended AdS-Schwarzschild geometry, with Kruskal-Szekeres coordinates $T,X$ and lightcone coordinates $U,V$ indicated. Of course, the singularity actually appears as a hyperbola in $T,X$. This diagram is a conformally-transformed sketch to indicate the general relationship among the coordinates; see \Ref{Fidkowski:2003nf} for more discussion.  Regions I through IV are defined by \Eq{eq:KS_coords}.}
\label{fig:Coordinate_Diagram}
\end{figure}

We consider the maximally extended AdS-Schwarzschild geometry \cite{griffiths2009exact,Lake:1977ui}, which, following \Ref{ER=EPR}, we will interpret as an Einstein-Rosen bridge connecting two black holes. The metric for the AdS-Schwarzschild black hole in $D$ spacetime dimensions is \cite{Hemming,Quasinormal}
\be
\mathrm{d}s^2 = -f(r)\mathrm{d}t^2 + \frac{\mathrm{d}r^2}{f(r)} + r^2 \mathrm{d}\Omega_{D-2}^2,\label{eq:metric}
\ee
where $\mathrm{d}\Omega_{D-2}^2$ is the surface element of the unit $(D-2)$-dimensional sphere and $f(r)$ is defined to be
\be
f(r) = 1 - \frac{16\pi G_D M}{(D-2)\Omega_{D-2} r^{D-3}} + \frac{r^2}{L^2}, 
\ee
writing $G_D$ for Newton's constant in $D$ dimensions, $\Omega_{D-2} = 2\pi^{(D-1)/2}/\Gamma[(D-1)/2]$ for the area of the unit $(D-2)$-sphere, and $L$ for the AdS scale. The horizon $r_{\rm H}$ is located at the point where $f(r_{\rm H}) = 0$. The tortoise coordinate can be defined as $r^* = \int \mathrm{d}r/f(r)$, the ingoing and outgoing Eddington-Finkelstein coordinates $v = t + r^*$ and $u = t - r^*$, with which we can define the lightcone Kruskal-Szekeres coordinates
\be 
\begin{aligned}
\text{(I)}&\;\;& U &= - e^{-f'(r_{\rm H})u/2} &\;\; V &=  e^{f'(r_{\rm H})v/2}\\
\text{(II)}&\;\;& U &=  e^{-f'(r_{\rm H})u/2} &\;\; V &=   e^{f'(r_{\rm H})v/2}\\
\text{(III)}&\;\;& U &=  e^{-f'(r_{\rm H})u/2} &\;\; V &= - e^{f'(r_{\rm H})v/2}\\
\text{(IV)}&\;\;& U &= - e^{-f'(r_{\rm H})u/2} &\;\; V &= - e^{f'(r_{\rm H})v/2}.
\end{aligned}\label{eq:KS_coords}
\ee
Regions I through IV are depicted in \Fig{fig:Coordinate_Diagram} and define the maximally extended AdS-Schwarzschild black hole geometry. Defining $T=(U+V)/2$ and $X=(V-U)/2$, the horizon is located at $T=\pm X$, that is, at $UV = 0$, while the singularity is located at $T^2 - X^2 = 1$. The one-sided AdS black hole occupies Region I and half of Region II, i.e., $V>0, X>0$. In these coordinates, the metric becomes
\be
\begin{aligned}
\mathrm{d}s^2 &= -\frac{4 |f(r)|e^{-f'(r_{\rm H})r^*}}{[f'(r_{\rm H})]^2}\mathrm{d}U\mathrm{d}V + r^2 \mathrm{d}\Omega_{D-2}^2\\
&=\frac{4 |f(r)|e^{-f'(r_{\rm H})r^*}}{[f'(r_{\rm H})]^2}(-\mathrm{d}T^2 + \mathrm{d}X^2) + r^2 \mathrm{d}\Omega_{D-2}^2,
\end{aligned}\label{eq:metricKruskal}
\ee
where $r$ is now defined implicitly in terms of $U$ and $V$ via
\be
UV = T^2 - X^2 =  \pm e^{f'(r_{\rm H})r^*},\label{eq:UV}
\ee
where the sign is $-$ for Regions I and III and $+$ for Regions II and IV.

\begin{figure}[t]
  \begin{center}
    \includegraphics[height=0.3\textwidth]{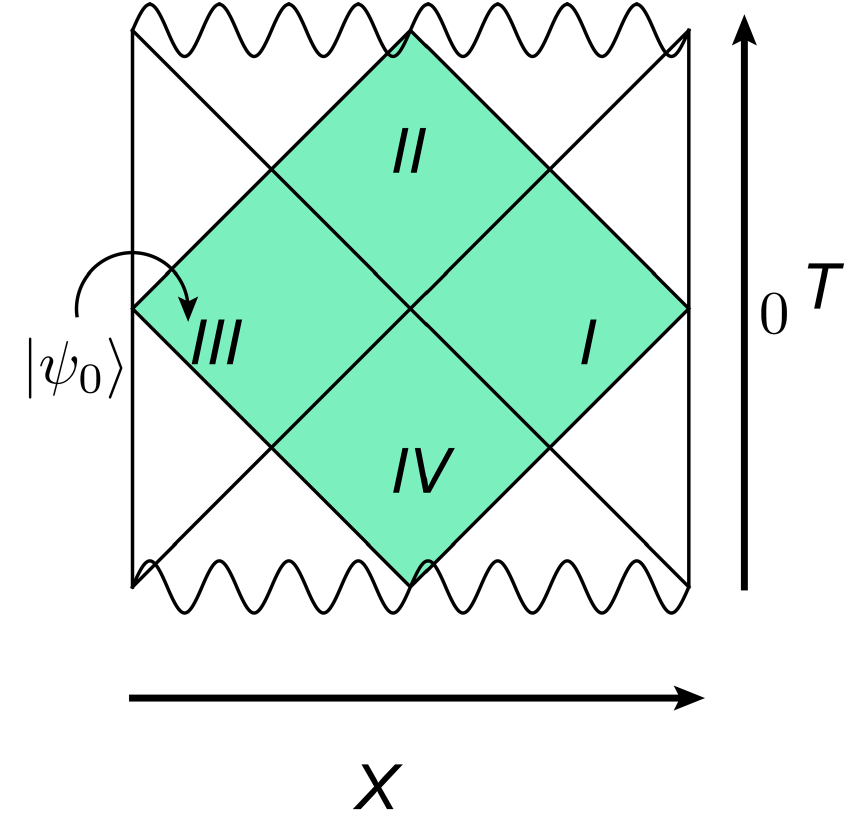}\qquad\qquad \includegraphics[height=0.3\textwidth]{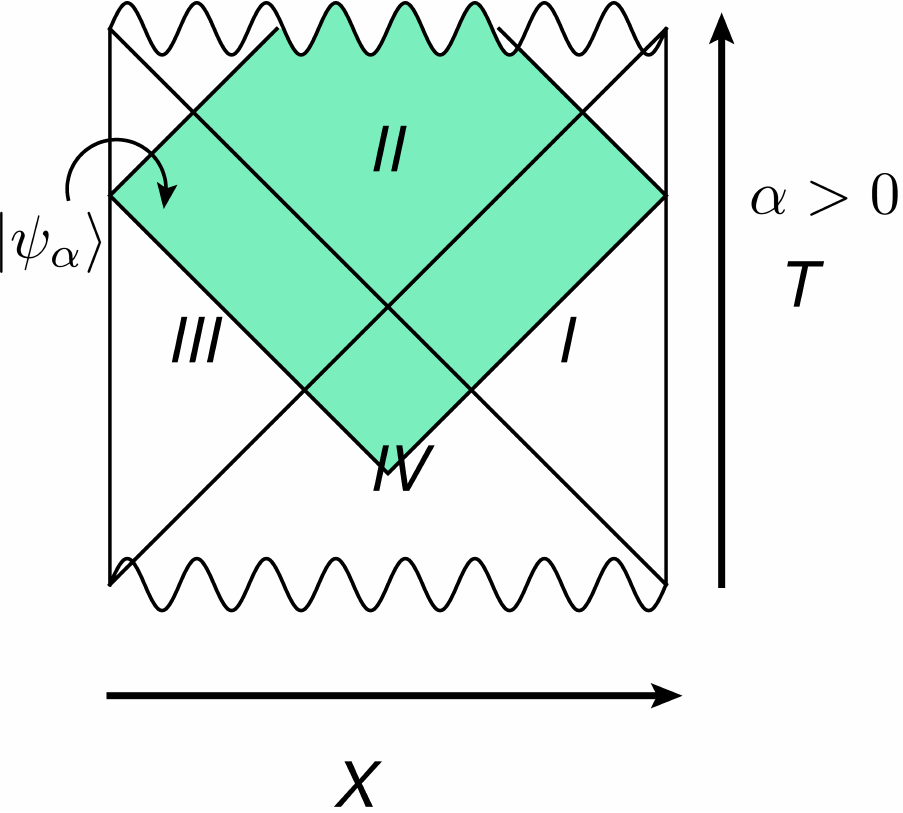}
  \end{center}
  \caption{(left) The state $\ket{\psi_0}$, corresponding to a wormhole geometry where the ER bridge intersects the boundary at $T=0$. (right) The family of states $\ket{\psi_\alpha}$, $\alpha>0$, for which the ER bridge intersects the boundary at $T>0$. 
  }
\label{fig:ER_Bridge}
\end{figure}

We now turn to the CFT interpretation of the geometry. In the Maldacena and Susskind proposal of ER=EPR \cite{ER=EPR}, it is pointed out that, in AdS/CFT, the state $\ket{\psi(t)}$ in the CFT corresponds at different times to different causal diamonds in the eternal, maximally extended AdS-Schwarzschild geometry. Different spatial slices through a given causal diamond that intersect the boundaries at fixed points are related to each other by the Wheeler-deWitt equation in the bulk. If one is outside a black hole in AdS, without knowing {\it a priori} which time slice one is on, then the different $\ket{\psi(t)}$ are simply a one-parameter family of states $\ket{\psi_\alpha}$, where $\alpha$ has replaced $t$ and is now just the label for the state of the CFT at time $t=0$; all of the $\ket{\psi_\alpha}$ describe pairs of black holes containing some sort of ER bridge. The various geometries are shown in \Fig {fig:ER_Bridge}.

Famously, the maximally extended AdS black hole can be described on the CFT side of the AdS/CFT correspondence by the thermofield double state of two noninteracting large-$N$ CFTs on the boundary sphere. We take the interpretation \cite{ER=EPR} of the state as two entangled black holes that both evolve forward in time, that is,
\be
\ket{\psi(t)} = \frac{1}{\sqrt{Z}}\sum_n e^{-\beta E_n/2}e^{-2iE_n t}\ket{\bar n}_L \otimes \ket{n}_R,\label{eq:thermofield}
\ee
where $\ket{n}_L$ and $\ket{n}_R$ are the $n^{\text{th}}$ eigenstates of the left and right CFTs, respectively, with eigenvalue $E_n$, a bar denotes the CPT conjugate, and $\beta$ is the inverse temperature.
We note that the CFT time $t$ in \Eq{eq:thermofield} is the $r\rightarrow \infty$ limit of the Schwarzschild time $t$ that appears in \Eq{eq:metric}. By considering the surface of constant Kruskal time $T$ that intersects the $r=\infty$ boundary at Schwarzschild time $t$, we can instead parameterize the CFT state corresponding to the eternal AdS black hole as $\ket{\psi(T)}$. Equivalently, we can write as $\ket{\psi_T}$ the family of ER bridges indexed by $T$, which correspond at the fixed Kruskal time $T=0$ to the CFT state $\ket{\psi(T)}$. The black hole described by $\ket{\psi_{T_0}}$ is given by the metric \eqref{eq:metricKruskal} with $T$ replaced with $T-T_0$ in \Eq{eq:UV}.
The analogous states with two one-sided black holes on the boundary CFTs will be called $\ket{\phi_T}$, where
\be
\ket{\phi_t} = \frac{1}{\sqrt{Z}}\left(\sum_m e^{-\beta E_m/2} e^{-iE_{m} t}\ket{\bar m}_L \right)\otimes \left(\sum_n e^{-\beta E_n/2} e^{-i E_{n} t} \ket{n}_R\right).
\ee

\section{The Single-Observer Case}

To gain intuition for the setup, in this section we restrict ourselves to measurements that a single (test particle) observer can perform in an otherwise empty (AdS-)Schwarzschild spacetime. Such observers are forbidden from receiving information from or coordinating with other observers; that is, we first investigate the aspects of the geometry that can be probed by a single causal geodesic. We will refer to this class of observers as \textit{isolated} observers. 
The simplest way for an isolated observer to verify the existence of an ER bridge would be to pass through it, i.e., to traverse the wormhole. It turns out, however, that this process is disallowed both by classical general relativity and, via ER=EPR, by quantum mechanics. 

In general relativity, the nontraversability of wormholes follows immediately from a more fundamental result, the topological censorship theorem \cite{Friedman:1993ty}, which is the statement that in a globally hyperbolic, asymptotically flat spacetime satisfying the null energy condition (NEC), any causal curve from past null infinity to future null infinity is diffeomorphic to an infinite causal curve in topologically trivial spacetime (such as Minkowski space). In other words, no causal observer's worldline can ever probe nontriviality of topology of spacetime.\footnote{Of course, nonisolated observers \textit{can} determine topological characteristics of their spacetime, for example by seeing the same stars on opposite sides of the sky and thereby determining that spatial sections of their spacetime are toroidal. However, they must receive information from outside their worldline---in this case, photons emitted by distant stars that travel on topologically distinct geodesics---to do so. Furthermore, the topological censorship theorem guarantees that if the spacetime is asymptotically flat, satisfies the NEC, and allows Cauchy evolution, then any handles must collapse to a singularity before an observer can travel around them.} Probing the nontrivial topology of an ER bridge simply means passing through the wormhole, which is therefore forbidden given the NEC.
In a previous paper \cite{BPR}, we showed that violation of the NEC in ER=EPR necessarily leads to violation of the no-cloning theorem and the breakdown of unitary evolution. Traversable wormholes are therefore also forbidden by quantum mechanics given ER=EPR, as they would correspond to a breakdown of unitarity by allowing superluminal signaling.

The next simplest means of verifying the existence of an ER bridge would be to detect the nontrivial topology of the wormhole without traversing it. In the present context, we see that detecting the nontrivial topology is equivalent under ER=EPR to detecting the existence of entanglement---more precisely, to constructing a linear operator that detects if an unknown state is entangled with anything else. But it is well known that such an operator is forbidden by the linearity of quantum mechanics, as \Ref{MP} discusses. Briefly, this is because projection operators cannot project onto a subspace unless that subspace is closed under superposition. An attempt to project onto the set of all entangled states will therefore fail due to the set of all entangled states not being closed under superposition; such a projector will inevitably project onto the entire Hilbert space of all states. On the gravity side, this leads to a result stronger than the nontraversability of wormholes: not only does ER=EPR forbid an observer from traversing wormholes, it forbids an isolated observer from verifying their existence even once inside them. 

This result can be straightforwardly verified in general relativity by examining the applicable metrics.  Importantly, the metric given in \Eqs{eq:metricKruskal}{eq:UV} for the maximally extended geometry has several isometries: it is invariant under the exchange $(U,V)\leftrightarrow(-U,-V)$ and also under the exchanges $(T,X)\leftrightarrow(T,-X)$ and $(T,X)\leftrightarrow(-T,X)$. That is, Regions I and II in \Fig{fig:Coordinate_Diagram} are the same as Regions III and IV, respectively, and moreover the entire metric is symmetric under spatial ($X$) or temporal ($T$) reversal. In particular, the regions present in both this geometry and the one-sided black hole geometry (Region I and half of Region II, i.e., $V>0, X>0$) are completely identical in the two cases. It is this property that implies that an observer on a geodesic entering a one-sided black hole cannot distinguish it from a two-sided black hole via any local measurement of curvature. 

We have therefore shown that a single (isolated) observer cannot observe whether a given black hole hosts an ER bridge, even by jumping into it. We next consider observables that require multiple communicating observers to implement.

\section{The Multiple-Observer Case}

\begin{figure}[t]
  \begin{center}
    \includegraphics[width=0.3\textwidth]{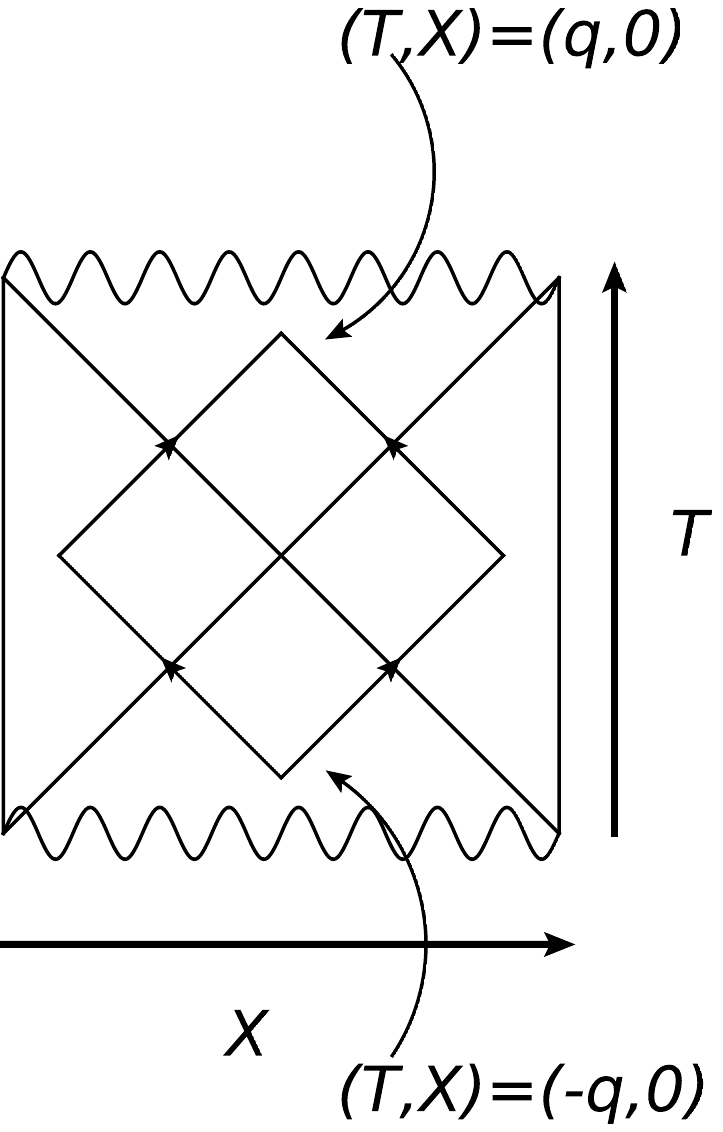}
  \end{center}
  \caption{The procedure described in the text for detecting a wormhole. Alice and Bob emerge from the white hole portion of the AdS-Schwarzschild geometry, then meet again inside the black hole.}
\label{fig:Alice_Bob_1}
\end{figure}

One can ask the question of whether two (or, for that matter, many) observers can detect the existence of entanglement or, equivalently, of nontrivial topology. The setup of the experiment is as follows. Consider a maximally extended, eternal AdS-Schwarzschild geometry, as depicted in \Fig{fig:Alice_Bob_1}. Allow two observers, Alice and Bob, to initially begin in the white hole portion of the geometry. (We will consider the case of more than two observers later in this section.) Now let the observers exit the white hole\footnote{We note that the white holes mentioned in our construction are for convenience only; it suffices for Alice and Bob to have communicated at some past time and simply to have moved out of causal contact. Indeed, it is possible for Alice and Bob to both exist in the same asymptotically AdS vacuum, as long as a wormhole exists connecting their locations. It is, however, necessary for them to enter the wormhole in order to attempt to detect information regarding the entanglement in this picture.} to the two different asymptotic regions not contained in the black hole. Next, they both jump into their respective black holes and compare notes. In such a way, they could potentially determine if there was entanglement before hitting the singularity.

\begin{figure}[t]
  \begin{center}
    \includegraphics[width=0.3\textwidth]{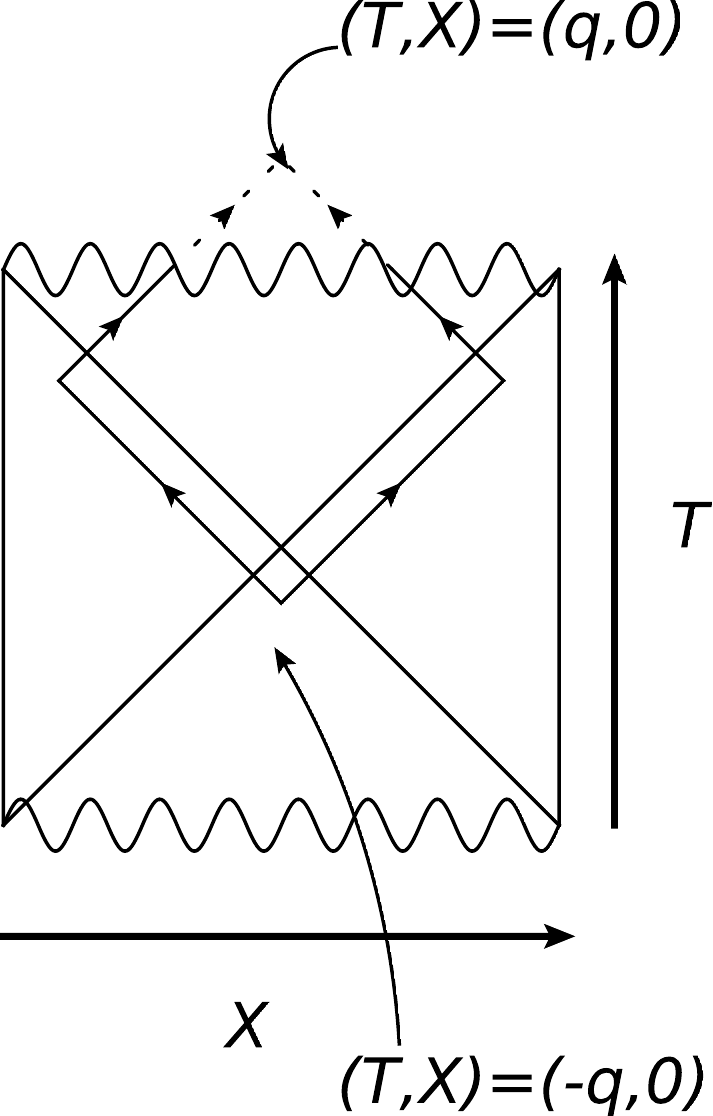}
  \end{center}
  \caption{Unlike in \Fig{fig:Alice_Bob_1} above, here the geometry is shifted to some $\ket{\psi_\alpha}$ for some sufficiently large $\alpha\neq 0$; Alice and Bob hit the singularity before they can meet and are therefore unable to verify the existence of a wormhole.}
\label{fig:Alice_Bob_2}
\end{figure}

The problem with this construction is that it doesn't definitively tell the observers if there was entanglement or not. Indeed, Alice and Bob could jump into the $\ket{\psi_0}$ ER bridge at sufficiently late time $T$ that they are unable to communicate (since one or both of them will hit the singularity before being able to do so); equivalently, the geometry could be $\ket{\psi_\alpha}$ for $\alpha$ too large (instead of $\ket{\psi_0}$), as depicted in \Fig{fig:Alice_Bob_2}. The same argument that states that no linear operator permits the observers from detecting whether or not there is entanglement precludes this verification procedure from succeeding with probability 1. But how is it possible to reconcile the fact that Alice and Bob can sometimes verify the existence of an ER bridge with the impossibility of projecting onto a generic family of states?

To be concrete, suppose that Alice and Bob are in some geometry in the set of all $\ket{\psi_\alpha}$ (for $\alpha$ unknown) and travel on outgoing nearly null trajectories beginning in the white hole at $(T,X)=(-q,0)$, with Alice (Bob) entering Region III (respectively, I) at $(T,X)\simeq(-q/\sqrt{2},\mp q/\sqrt{2})$, turning around at $(T,X) \simeq (0,\mp q)$, and entering Region II through their respective black holes at $(T,X)\simeq(q/\sqrt{2},\mp q/\sqrt{2})$. Now, if they do not hit a singularity, their geodesics will cross again at $(T,X) \simeq (q,0)$. That is, their geodesics will cross if they are in the state $\ket{\psi_\alpha}$, i.e., the state in which $T$ is shifted by $\alpha$, for $\alpha < 1-q$. If $\alpha>1-q$ for a $\ket{\psi_\alpha}$ state or if they had instead been in any one of the $\ket{\phi_\alpha}$ states, they would hit the singularity without their paths ever crossing. (Recall that for $X=0$, the singularity is located at $T=\pm 1$ for the $\ket{\psi_0}$ geometry.) Hence, Alice and Bob are able to verify if they are in the set $S_q=\{\ket{\psi_\alpha}|\alpha<1-q\}$. 

However, this thought experiment does not require the existence of a projection operator onto the entire family $S_q$. Instead, after their geodesics cross, Alice and Bob can actually determine in which of the $\ket{\psi_\alpha}$ they are.  All null geodesics from the horizon to the singularity are isomorphic and experience the same pattern of values of the curvature tensor on the way in. That is, a family of null geodesics with, e.g., constant $V = V_0$ can be labeled by the time $t$ at which they cross a surface at fixed proper distance from the horizon in Region I, which is the only difference among the geodesics; since the metric \eqref{eq:metric} is independent of $t$, all of these geodesics experience the same inward journey. Hence, before meeting Bob inside Region II, there is no distinguishing event by which Alice can measure $\alpha$. However, the value of the Riemann tensor {\it at the moment Alice's and Bob's geodesics cross} is unique for each $\ket{\psi_\alpha}$. 

In particular, at the moment their geodesics cross, Alice and Bob can measure the tidal forces acting in their local Lorentz frames by computing some component of the Weyl tensor. At $(T,X)=(q,0)$, \Eq{eq:UV} implies that, in Region II, $r^*$ and hence $r$ is a monotonic function of $(q-\alpha)$. (Recall that for $\ket{\psi_\alpha}$ all equations describing the metric are shifted by $T\rightarrow T-\alpha$.) Let us define a local Lorentz frame in coordinates $(\hat t, \hat r, \hat \theta)$, where $\hat \theta$ is the orthonormal coordinate in the $D-2$ angular directions. When their paths cross, Alice and Bob can measure the $\hat r \hat \theta \hat r \hat \theta$ component of the Weyl tensor, which is
\be 
W_{\hat r \hat \theta \hat r \hat \theta} = -\frac{1}{L^2} - \left(\frac{D-3}{D-2}\right)\frac{8\pi G_D M}{\Omega_{D-2}r^{D-1}}.
\ee
Note that this quantity monotonically increases as $r$ [and so in $(q-\alpha)$]. This implies that Alice and Bob can determine $\alpha$ by measuring tidal forces at the moment when their geodesics cross; there is a bijection between $\alpha$ and the size of the tidal force. This measurement thus acts as a projection operator $P_\alpha = \ket{\psi_\alpha}\bra{\psi_\alpha}$. This is analogous to the possibility of being able to detect if two qubits are in some particular entangled state, rather than absolutely any entangled state whatsoever. 

The key point here is that if the observers hit the singularity before exchanging a signal, i.e., if the wavefunction is one of the $\ket{\psi_\alpha}$ for which $\alpha>1-q$, then Alice and Bob are unable to confirm the existence of the ER bridge. If $\alpha<1-q$, the experiment Alice and Bob perform actually determines $\alpha$. This procedure therefore fails to determine if the region behind a horizon contains a generic wormhole: it can sometimes reveal its existence, but not rule out its presence. It therefore does not implement a projector onto the set of all wormhole states. Thus, no contradiction with linearity of quantum mechanics arises in ER=EPR from the ability of Alice and Bob to jointly explore the wormhole geometry.

{\it A priori}, one could wonder whether even more general configurations of more than two observers could make the existence of wormhole topology into an observable. Note that it is not consistent to consider a setup in which there is an infinite set of observers (or  signals) entering a horizon at earlier and earlier times, as this would violate the necessary assumption of weak backreaction and hence invalidate the AdS-Schwarzschild spacetime ansatz. Hence, in a given slicing of spacetime, there must be an initial observer to enter the horizon. A prototypical setup for the thought experiment with more than two observers can therefore be rephrased as follows. After meeting and arranging the experiment, Bob and Alice go their separate ways. Bob jumps into his horizon, crossing it at spacetime point $p=(T,X)\simeq(q,q)/\sqrt{2}$ as before. This time, however, Alice remains outside her horizon and instead sends into her black hole multiple light pulses at regular intervals, with the first light pulse she emits (after leaving Bob) entering her wormhole mouth at $p'=(T,X)\simeq(q',-q')/\sqrt{2}$.   The multiple light pulses are equivalent to having multiple observers enter the black hole at different times. However, one can choose a slicing of spacetime in which $p$ and $p'$ are on the same spacelike sheet; that is, one can simply apply a boost to equate the spatial components of $p$ and $p'$. Since a boost can be independently applied to each asymptotically AdS spacetime, it follows that the case in which Bob is also replaced by multiple observers can be similarly simplified. As a result, the multiple observer setup reduces to the two observer setup, which we showed previously cannot definitively answer the question of whether there is a wormhole geometry. 
 
Thus, even with multiple observers, the measurement of whether or not there is an ER bridge in general is not a valid observable, any more than the question of whether two qubits are arbitrarily entangled is a quantum mechanical observable.

\section{Conclusions}

The ER=EPR proposal is a compelling but surprising idea about quantum gravity, identifying features of ordinary quantum mechanics with geometrical and topological features of spacetime. As an extraordinary claim, it is necessary that it be subjected to rigorous theoretical tests to ascertain whether it suffers from any inconsistencies. One such potential issue, which we have addressed in this paper, is whether ER=EPR implies a serious modification of quantum mechanics, namely, the introduction of state dependence. The argument that ER=EPR implies state dependence rests on the observation that the correspondence identifies entanglement with wormholes. Famously, entanglement is not a quantum mechanical observable, so this leads to the question of whether the observation of a wormhole contradicts, under ER=EPR, linearity of quantum mechanics. 

In this paper, we have argued that ER=EPR does not contradict this principle of quantum mechanics precisely because the general question of the existence or nonexistence of a wormhole is also not an observable. We showed that neither a single observer nor a group of observers is able to definitively establish whether a pair of event horizons is linked by an ER bridge. A single observer can never detect the (nontraversable) wormhole's existence, which mirrors the fact that, given a single qubit, one cannot tell if it is entangled by anything else. On the other hand, by exploring the spacetime, two or more observers working in concert can decide if they are in a particular ER bridge geometry, but cannot project onto the entire family. Under ER=EPR, this statement mirrors the fact that one can project two qubits onto a particular entangled state but not onto the family of all possible entangled states.

Many options are available for future investigation. The ER=EPR correspondence has been subjected to some tests \cite{BPR,JensenKarch,Sonner,Baez:2014bka,Gharibyan:2013aha,Mandal:2014wfa}, but the challenge of seeing the duality between wormholes and any arbitrary form of quantum entanglement remains, as does the very definition of what is meant by a ``wormhole'' in ER=EPR for theories without a weakly-coupled holographic gravity dual. Other open issues include the investigation of whether firewalls are truly nongeneric in ER=EPR \cite{ER} and whether the correspondence can be concretely realized outside of asymptotically AdS spacetime. The answers to these questions and others will likely provide important insight in future investigations in the connections between entanglement and spacetime geometry.

\begin{center} 
 {\bf Acknowledgments}
 \end{center}
 \noindent 

We thank Sean Carroll and Clifford Cheung for helpful discussions.
This research was supported in part by DOE grant DE-SC0011632 and by the Gordon and Betty Moore Foundation through Grant 776 to the Caltech Moore Center for Theoretical Cosmology and Physics. N.B. is supported by the DuBridge postdoctoral fellowship at the Walter Burke Institute for Theoretical Physics. G.N.R. is supported by a Hertz Graduate Fellowship and a NSF Graduate Research Fellowship under Grant No.~DGE-1144469.

\bibliographystyle{utphys}

\bibliography{EPRStateBib}

\end{document}